\documentclass[aps,reprint]{revtex4-1}
\usepackage[english]{babel}
\usepackage{graphicx}
\usepackage{dcolumn}
\usepackage{bm}
\usepackage{amssymb}
\usepackage{amsmath}
\usepackage{wasysym}
\usepackage{chngcntr}
\usepackage{textcomp}
\usepackage{hyperref}
\usepackage{setspace}

\renewcommand*{\thetable}{\arabic{table}}
\renewcommand*{\thefigure}{\arabic{figure}}
\renewcommand*{\theequation}{\arabic{equation}}

\begin{document}
\title{Electrical N\'{e}el-order switching in magnetron-sputtered CuMnAs thin films}
\author{T.~Matalla-Wagner}
\email{tristan@physik.uni-bielefeld.de}
\author{M.-F.~Rath}
\author{D.~Graulich}
\author{J.-M.~Schmalhorst}
\author{G.~Reiss}
\author{M.~Meinert}
\email{meinert@physik.uni-bielefeld.de}
\affiliation{Center for Spinelectronic Materials and Devices, Bielefeld University, Universit\"atsstra\ss e 25, D-33501 Bielefeld, Germany}

\date{\today}

\keywords{}

\begin{abstract}
Antiferromagnetic materials as active components in spintronic devices promise insensitivity against external magnetic fields, the absence of own magnetic stray fields, and ultrafast dynamics at the picosecond time scale. Materials with certain crystal-symmetry show an intrinsic N\'{e}el-order spin-orbit torque that can efficiently switch the magnetic order of an antiferromagnet. The tetragonal variant of CuMnAs was shown to be electrically switchable by this intrinsic spin-orbit effect and its use in memory cells with memristive properties has been recently demonstrated for high-quality films grown with molecular beam epitaxy. Here, we demonstrate that the magnetic order of magnetron-sputtered CuMnAs films can also be manipulated by electrical current pulses.  The switching efficiency and relaxation as a function of temperature, current density, and pulse width can be described by a thermal-activation model. Our findings demonstrate that CuMnAs can be fabricated with an industry-compatible deposition technique, which will accelerate the development cycle of devices based on this remarkable material.
\end{abstract}

\maketitle

\counterwithout{equation}{section}

\section{Introduction}

\begin{figure*}[t]
	\centering
	\includegraphics{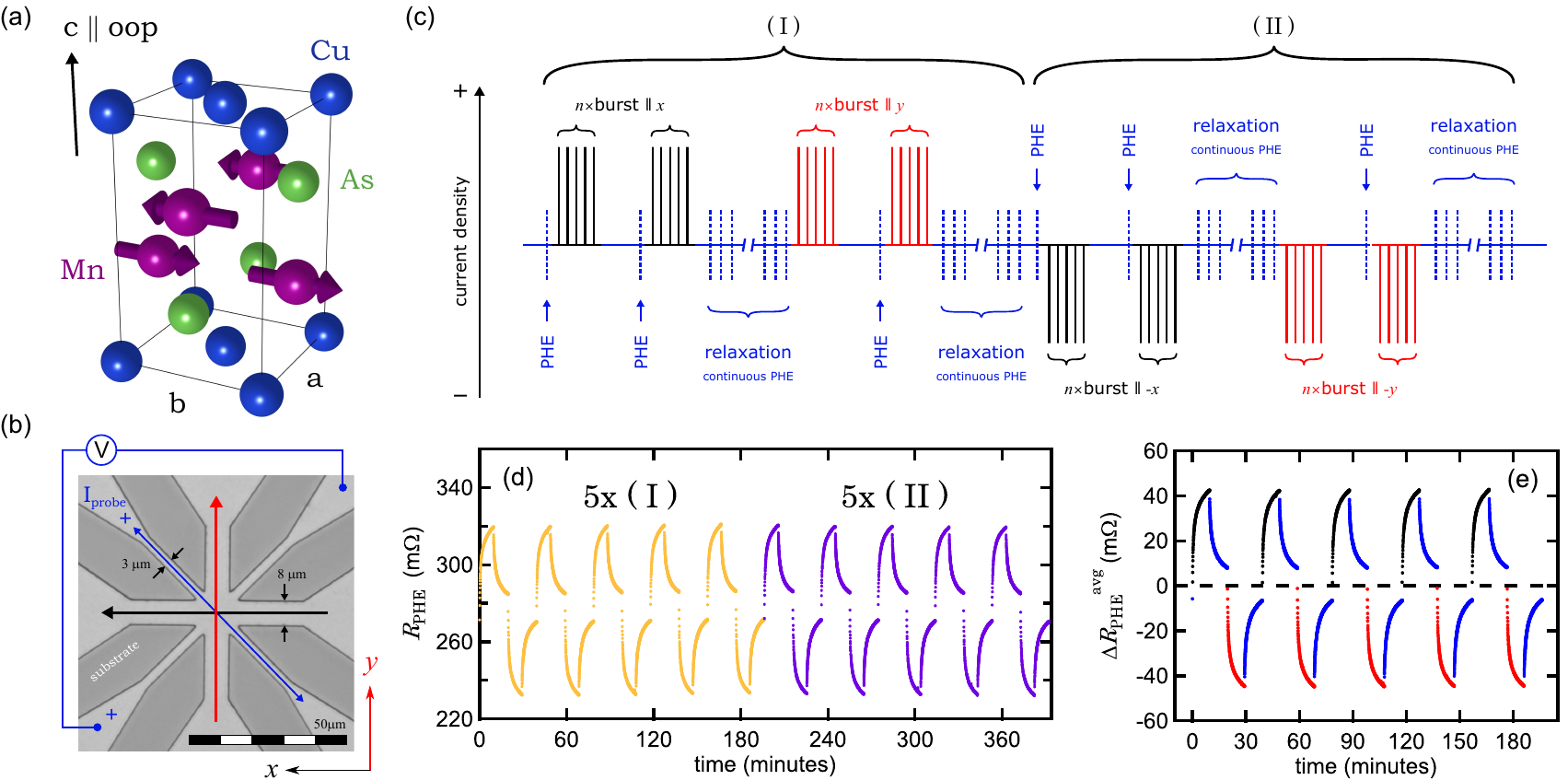}
	\caption[fig1]{\textbf{Device and measurement scheme.}
		(a)~Unit cell of tetragonal CuMnAs with antiferromagnetically coupled Mn moments that form structural inversion partners. The $c$-axis points out-of-plane.
		(b)~Greyscale optical micrograph of the patterned film. Current pulses can be applied to the pulse lines parallel to the $x$- or $y$-axis. The arrows indicate the conventional current direction for positive polarity.
		(c)~Sequence of bursts and $R_\text{PHE}$ measurements with color code as in (a). Part (\,I\,) and (\,II\,) are identical but with inverted polarity of the current pulses. The current density used to probe $\bm{L}$ is two orders of magnitude smaller than used for switching. A duty cycle of $10^{-3}$ separates the individual pulses within a burst.
		(d)~$R_\text{PHE}$ versus time trace at $T_\text{s}=260\,\text{K}$, $j = 5.9 \times 10^{10}\,\text{A}/\text{m}^2$ and $\Delta t = 5\,\text{\textmu s}$. Different polarity of the current pulses in part (\,I\,) and (\,II\,) is highlighted by different colors.
		(e)~Average over (\,I\,) and (\,II\,)  plotted in colors matched to the measurement scheme in (c). A constant offset in $\Delta R_\text{PHE}$ is removed.
		}
	\label{fig:1}
\end{figure*}

Observation and manipulation of the magnetic order in antiferromagnetic materials, also known as the N\'eel-order, is notoriously difficult because of their lack of a net magnetization. Likewise, information encoded in the antiferromagnetic state would be very well protected against external influences once it could be written. Thanks to ultrafast magnetization dynamics of antiferromagnets, the magnetic state can---in principle---be manipulated on a ps time scale. Unfortunately, only few mechanisms for the manipulation of antiferromagnetic states are known, such as exchange coupling to a ferromagnet, magnetoelastic coupling, and the spin-flop transition. More recently, advances have been made in exploiting multiferroic coupling, which is, however, only available in insulating materials~\cite{Song2018}. A few years ago, the N\'{e}el-order spin-orbit torque (NSOT) was predicted by \v{Z}elezn\'{y} \textit{et al.} as a mechanism to manipulate the N\'{e}el-order in tetragonal Mn$_2$Au~\cite{Zelezny2014}. It generally occurs in antiferromagnets where the magnetic sublattices A and B with magnetic moments $\bm{m}_\mathrm{A} = - \bm{m}_\mathrm{B}$ are connected via structural inversion, i.e. the material exhibits combined \textit{PT} (parity and time) symmetry. An electrical current-density $\bm{j}$ flowing through an antiferromagnet with this symmetry gives rise to the inverse spin-galvanic effect, which generates local spin accumulations and eventually results in a torque acting on the N\'{e}el-vector $\bm{L} = \bm{m}_\mathrm{A} - \bm{m}_\mathrm{B}$ favoring an orientation $\bm{L} \perp \bm{j}$. Current-induced N\'{e}el-order switching has been observed by measuring the planar Hall resistance $R_\text{PHE}$ in tetragonal CuMnAs~\cite{Wadley2016,Olejnik2017} fabricated by molecular beam epitaxy (MBE) and in tetragonal Mn$_2$Au~\cite{Bodnar2018,Meinert2018,Zhou2018} fabricated by sputtering. However, the proposed NSOT mechanism alone does not explain the strong dependencies of the switching efficiency on sample temperature and current density. A macroscopic thermal-activation model proposed by some of the authors of this study extends the description by taking the NSOT as an effective field acting on $\bm{L}$ with a switching energy-barrier $E_\mathrm{B}$ that can be overcome by thermal activation~\cite{Meinert2018}. Here we show switching experiments on CuMnAs films deposited by magnetron sputtering. We investigate the dependence of the $R_\text{PHE}$ change on the sample temperature, the pulse current-density, and the pulse width. The results are discussed with the help of our thermal activation model, and we show that many aspects of the observed behavior are described well by the model on a semi-quantitative level.

\section{Experimental details}

Samples of tetragonal \mbox{Cu}\mbox{Mn}\mbox{As} (cf.~Fig.\ref{fig:1}\,(a)) are grown in stacks of the type GaAs~/ \mbox{Cu}\mbox{Mn}\mbox{As}~68\,nm~/ Ti~5\,nm using dc~magnetron sputtering (cf. Appx.~\ref{Properties})). The film is patterned in star-like structures, shown in Fig.\,\ref{fig:1}\,(b). We apply current bursts consisting of
\begin{align}\label{eq:PulsesPerBurst}
N(\Delta t, j) = Q \, \left( \Delta t \, j \, w \, h \right)^{-1}
\end{align}
pulses with pulse width $\Delta t$ and current density $j = | \bm{j} |$ to the pulse lines of our device. An arbitrary waveform generator (Agilent 33522A) in combination with a differential broadband amplifier (Tabor Electronics 9260) are used as voltage source. The charge per burst $Q = 10\,\text{\textmu C}$ is kept constant in all experiments. $w$ and $h$ are the current line width and film thickness, respectively. All measurements are performed in a closed-cycle He cryostat allowing for a variation of the sample temperature $T_\text{s}$. The pulse voltage $V_{x,y}(T_\text{s}) = j \, w h \, R_{x,y}(T_\text{s})$ is adjusted so that $j$ stays the same for both current lines at all temperatures by measuring $R_{x,y}$ before each sequence. Switching between pulse and probe lines is done with reed relays. The measurement sequence is sketched in Fig.~\ref{fig:1}\,(c). Before the first and after each burst, $R_\text{PHE}$ is measured with a Zurich Instruments MFLI lock-in amplifier at a frequency $f=81.3\,\text{Hz}$. Between burst and $R_\text{PHE}$-measurement we set a delay of $2\,\text{s}$. After $n=200$ bursts in $x$-direction, the relaxation of $R_\text{PHE}$ is observed over $T_\text{relax} = 600\,\text{s}$ with measurements every second. The same scheme is executed for bursts parallel to the $y$-direction, forming part (\,I\,) of the sequence. Part (\,II\,) is a repetition of (\,I\,) with inverted polarity of the current pulses. A complete experiment for a given set of parameters consists of five full repeats of each part.

\begin{figure}
	\centering
	\includegraphics{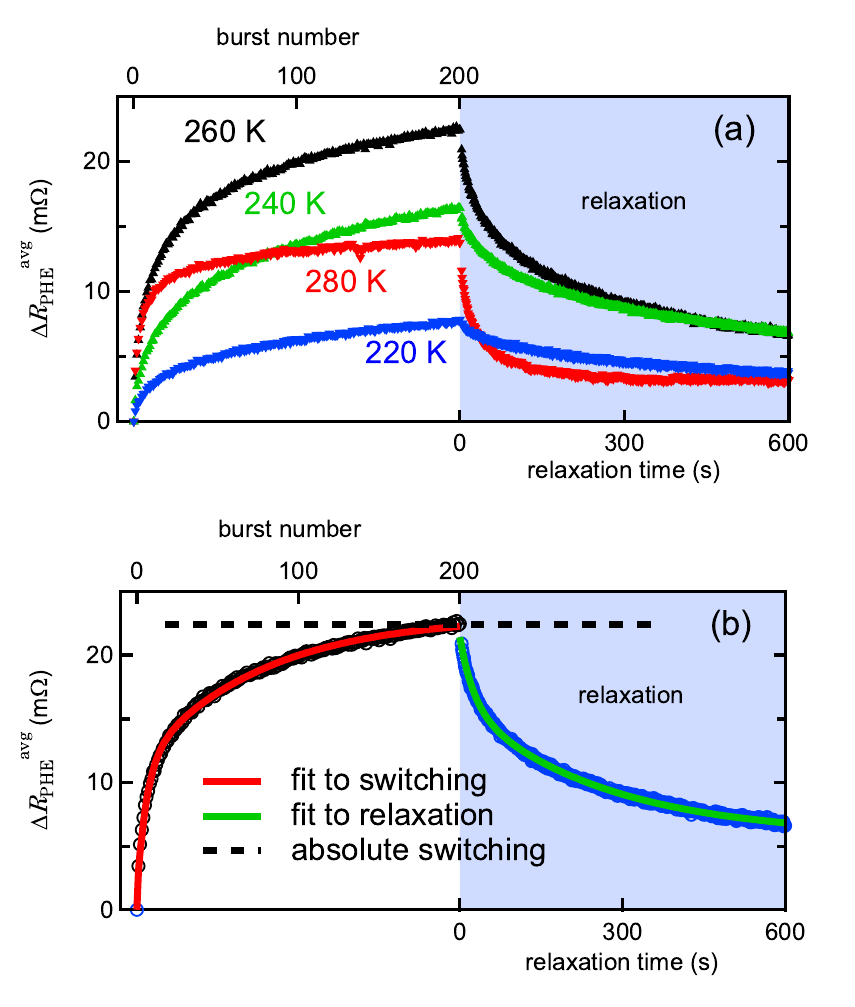}
	\caption[fig2]{\textbf{Shape and quantification of the switching.}
		First switching cycle for pulses $\parallel x$ using $j = 6.47 \times 10^{10}\,\text{A}/\text{m}^2$ and $\Delta t = 1\,\text{\textmu s}$. The curves are shifted to start at $\Delta R_\text{PHE}^\text{avg} = 0$ individually. Pulsing and relaxation are plotted on separate $x$-axis. (a) At different $T_\text{s}$. (b)~Plot of the data taken at $T_\text{s} = 260\,\text{K}$ in (a) using the colorcode introduced in Fig.~\ref{fig:1}\,(c). The solid lines in red and green are fits using Eq.~\eqref{eq:pulsing_fit} and ~\eqref{eq:relaxation_fit}, respectively.
	}
	\label{fig:2}
\end{figure}

Experiments are performed with ranges of sample temperatures $T_\text{s}$, current densities $j$, and pulse widths $\Delta t$. Exemplary data is shown in Fig.~\ref{fig:1}\,(d) where $R_\text{PHE}$ is drawn as a function of the elapsed time. Alternating the direction of pulsing switches the system between high and low values of $R_\text{PHE}$ where the change is independent of the current-polarity as expected for the NSOT mechanism. During $T_\text{relax}$ we see a decay of $\Delta R_\text{PHE}$. In Fig.~\ref{fig:1}\,(e) the average over part (\,I\,) and (\,II\,) is taken and an offset arising from imperfect lithography has been removed. This polarity-independent component of the signal shows perfectly reproducible switching of the planar Hall resistance with a slight asymmetry with respect to the pulsing direction. The shape of the switching is similar for different sets of parameters. The first switching cycle of a  $T_\text{s}$ variation is shown in Fig.~\ref{fig:2}\,(a). The temperature not only affects the amplitude of the switching curve, but also changes the overall shape, e.g. the steepness. In the following we discuss a procedure to quantitatively characterize the switching using only a few characteristic parameters. The observed switching behavior can be fitted using a sum of exponential functions. As seen in Fig.~\ref{fig:2}\,(b) three terms of the form
\begin{align}\label{eq:pulsing_fit}
R_\text{p}(b) = c_0 + c_1 \, \exp \left( - \frac{b}{\mu_1} \right)  + c_2 \, \exp \left( - \frac{b}{\mu_2} \right) 
\end{align}
during pulsing and
\begin{align}\label{eq:relaxation_fit}
R_\text{r}(t) = d_0 + d_1 \, \exp \left( - \frac{t}{\tau_1}\right) + d_2 \, \exp \left( - \frac{t}{\tau_2} \right)
\end{align}
during relaxation are sufficient to describe the observed switching curve reasonably well. $c_{0,1,2}$, $d_{0,1,2}$, $\mu_{1,2}$, and $\tau_{1,2}$ are fitting parameters where $d_0 \approx d_0 \, \exp \left( - t / \tau_0 \right)$ corresponds to an exponential function with $\tau_0 \gg T_\text{relax}$. $c_0$ denotes the saturation value of $\Delta R_\text{PHE}^\text{avg}$. Eq.~\eqref{eq:pulsing_fit} and \eqref{eq:relaxation_fit} are fitted piecewise to the averaged data. For the pulsing phase, we define the switching efficiency of the first burst $R_\mathrm{e}$ by taking the derivative of Eq.~\eqref{eq:pulsing_fit} at $b=0$:
\begin{align}\label{eq:SwitchingEfficiency}
R_\text{e} = 
\left| 
\frac{d R_\text{p}(b)}{db}
\right|_{b=0} 
= \left| \frac{c_1}{\mu_1} + \frac{c_2}{\mu_2} \right|
\end{align}
The decay is characterized by $\tau_{1,2}$ for a given set of parameters. We found that $c_1 / c_2 \approx d_1 / d_2$ for a given data set and, thus, this equality is enforced to improve the stability of the fit. As we will show later,  this constraint results naturally from the underlying ensemble physics. $R_\text{e}$ and $\tau_{1,2}$ can be calculated for each switching cycle where the signal-to-noise ratio is sufficiently large to apply the fit. We additionally define the difference of $R_\text{PHE}$ before and after applying $n$ bursts along one axis as the absolute switching amplitude $\left|\Delta R_\text{a}\right|$.

\section{Results}

\begin{figure*}
	\centering
	\includegraphics{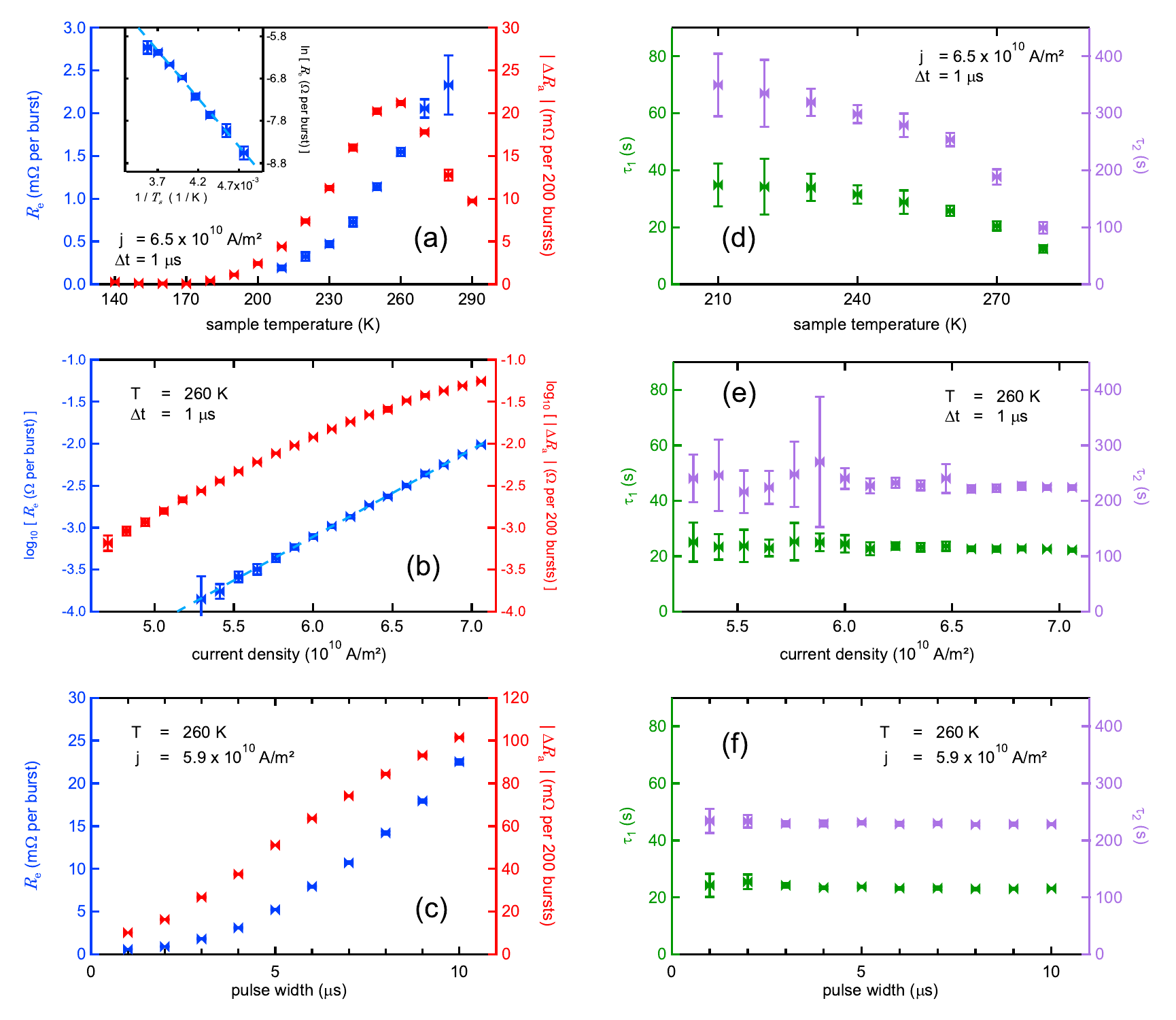}
	\caption[fig3]{\textbf{Fitting results of the parameter sweeps.}
		Dependences of $R_\text{e}$, $\left| \Delta R_\text{a} \right|$~(a\,-\,c) and $\tau_{1,2}$~(d\,-\,f) on $T_\text{s}$, $j$ and $\Delta t$.
		(a)~The inset shows an Arrhenius plot of $ R_\text{e} $ including a line fit disregarding the data point at $1/280\,\text{K}^{-1}$.
		(b)~Logarithmic representation of $R_\text{e}$ and $\left| \Delta R_\text{a} \right|$. The dashed line is a linear fit.
	}
	\label{fig:3}
\end{figure*}

The extracted characteristics $R_\text{e}$ and $\left|\Delta R_\text{a}\right|$ for different parameter sets $\left\{ T_\text{s}, j, \Delta t \right\}$ are shown in Fig.~\ref{fig:3}. In Fig.~\ref{fig:3}\,(a) we observe that $\left| \Delta R_\text{a} \right|$ has a local maximum at $T_\text{s}=260\,\text{K}$ and it decreases to zero for low $T_\text{s}$. $R_\text{e}$ monotonically increases with increasing $T_\text{s}$. The dependence of both characteristics on the current density is very strong, therefore we chose a logarithmic scale in Fig.~\ref{fig:3}\,(b). Line fits to the logarithmic data indicate an exponential dependence of $R_\text{e}$ on $j$. $\left| \Delta R_\text{a} \right|$ increases strongly with increasing $j$ as well, but deviates from an exponential law at high current density. $R_\text{e}$ and $\left| \Delta R_\text{a} \right|$ also increase with increasing $\Delta t$, as presented in Fig.~\ref{fig:3}\,(c). The relaxation time constants $\tau_{1,2}$ decrease with increasing $T_\text{s}$ and tend to zero for $T_\text{s}>280\,\text{K}$ as shown in Fig.~\ref{fig:3}\,(d). In contrast, $\tau_{1,2}$ are independent of $j$ and $\Delta t$ with $\tau_1 \approx 24\,\text{s}$ and $\tau_2 \approx 234\,\text{s}$ at $T_\text{s}=260\,\text{K}$ (cf. Fig.~\ref{fig:3}\,(e)\,\&\,(f)).
In Appx.~\ref{gap} we show further results of our analysis regarding the long-term stable $R_\text{PHE}$ change.

We record 10 switching cycles per experiment which are analyzed independently. Fig.~\ref{fig:3} shows the mean values. The results are slightly different between pulsing along $x$ or $y$. Hence, standard deviations are calculated separate for each direction and the greater value is taken as error to reflect the reproducibility of the switching.

\section{Discussion}

The presented experiments demonstrate that the switching of the N\'{e}el-vector $\bm{L}$ in our samples strongly depends on the sample temperature $T_\text{s}$, the current density $j$, and the pulse width $\Delta t$ although the total charge transfer per burst $Q$ is kept constant in each measurement. This finding is similar to our previous observations on Mn$_2$Au~\cite{Meinert2018} and is not explainable using solely the N\'{e}el-order spin-orbit torque (NSOT) mechanism for deterministic switching~\cite{Zelezny2014}. In the following, we use the macroscopic thermal-activation model \cite{Meinert2018} to interpret our data. We assume a film consisting of noninteracting grains of volume $V_\text{g}$ and $\bm{L}$ aligned along the $\pm x$- or $\pm y$-direction (cf. Fig.~\ref{fig:1}\,(b)). Without electrical current, all four states have equal energy but are separated by an energy barrier $E_\text{B} = K_{4\parallel}\,V_\text{g}$ with the in-plane biaxial magnetocrystalline anisotropy energy density $K_{4\parallel}$. The rate at which $\bm{L}$ attempts to change its orientation is given by the  N\'{e}el-Arrhenius equation
\begin{align}\label{eq:NeelArrhenius}
\frac{1}{\tau} = f_0 \, \exp \left( - \frac{E_\text{B}}{k_\text{B} T} \right)
\end{align}
with the Boltzmann constant $k_\text{B}$, the absolute temperature $T$ and the attempt rate $f_0$. We chose $f_0 = 10^{12}\,\mathrm{s}^{-1}$ to account for the THz dynamics in antiferromagnets \cite{Gomonay2016}. With each attempt, $\bm{L}$ can end up in four different states including the initial one. Using the rate equations (cf. Appx.~\ref{rate}), one can derive the time evolution of the planar Hall resistance as
\begin{align}\label{eq:RelaxationModel}
R_\text{PHE}(t) = R_\text{PHE}(t=0) \, \exp \left( - \frac{t}{\tau} \right)\text{.}
\end{align}
Details of the derivation are given in the Supplemental Material. Hence, the relaxation rates $\tau_i$ obtained from the experiment are linked to energy barriers via
\begin{align}\label{eq:EnergyBarrier}
E_\text{B}^i= \ln \left( f_0 \tau_i \right)\,k_B T_\text{s}\text{.}
\end{align}
We use Eq.~\eqref{eq:relaxation_fit} to fit the data and therefore obtain two distinct $E_\text{B}^{1,2}$ with relative occurrences $d_{1,2} / (d_1 + d_2)$. $E_\text{B}^i = K_{4\parallel}\,V_\text{g}^i$ depends on the grain size distribution in our sample which is unknown. However, we can represent the distribution of the grains that are active in our experiment by two slightly different grain sizes according to the two values of $E_\text{B}^{1,2}$. The relative occurrences of $d_i$ and $c_i$ depend on the fraction of the grains that are active at a particular parameter set $\left\{ T_\text{s}, j, \Delta t \right\}$. Thus, we enforce the ratios $d_1/d_2 = c_1/c_2$ to be constant for a given parameter set. With current flowing through the structure, the orientation of $\bm{L}$ perpendicular to the current becomes energetically favored due to the NSOT. The thermal-activation model states that for small changes (linear response) one can write \cite{Meinert2018}
\begin{align}\label{eq:LinearApproximation}
\left| \Delta R_\text{PHE}^\text{lin} \right| \approx A \Delta t f_0 \, \exp\left( \frac{L \chi V_\text{g} j}{\sqrt{2} k_\text{B} T V_\text{c}} - \frac{E_\text{B}}{k_\text{B} T}\right)
\end{align}
as the change of $R_\text{PHE}$ per pulse, where we assume switching of $ \bm{L} \parallel \bm{j} $ to $\bm{L} \perp \bm{j}$. Here, $L = |\bm{L}|$ denotes the N\'{e}el-vector magnitude, $\chi$ the spin-orbit torque efficiency (effective field per unit current density), $V_\text{c}$ the unit cell volume, and $A$ the effect amplitude. $A$ is equal to the anisotropic magnetoresistance (AMR) amplitude in the case of coherent switching of all grains. In our experiments we find $\Delta \rho = \left|\Delta R_\text{a}\right| \cdot 68\,\text{nm}$ up to $0.68\,\text{\textmu}\Omega\,\text{cm}$. Therefore, the AMR effect size has to be larger than $\Delta \rho / \rho = 0.14\%$. A burst of $N$ pulses results in a linear response of
\begin{align}\label{eq:BurstApproximation}
\left| \Delta R_\text{PHE}^\text{burst} \right| \approx \frac{Q A f_0}{j w h} \exp\left( \frac{L \chi V_\text{g} j}{\sqrt{2} k_\text{B} T V_\text{c}} - \frac{E_\text{B}}{k_\text{B} T}\right).
\end{align}
The relaxation between pulses is neglected and $N$ is calculated using Eq.~\eqref{eq:PulsesPerBurst}. $R_\text{e}$ is a measure for the switching at the first burst after relaxation, which is as close as we can get to the linear response case with approximately equal occupation of the four biaxial states. Therefore, we can write $ R_\text{e} \approx \left| \Delta R_\text{PHE}^\text{burst} \right| $. Thus, the model predicts a temperature dependence $ \ln R_\text{e} = s_1 - s_2/T $ and a dependence on the current density $\ln R_\text{e} =  m_1 j - m_2 - \ln j$, with constants $s_{1,2}$, and $m_{1,2}$. With this pulsing scheme, $R_\text{e}$ does not explicitly depend on $\Delta t$; however, the pulse width enters implicitly through the film temperature, which is not constant due to Joule heating.

\begin{figure}
	\centering
	\includegraphics{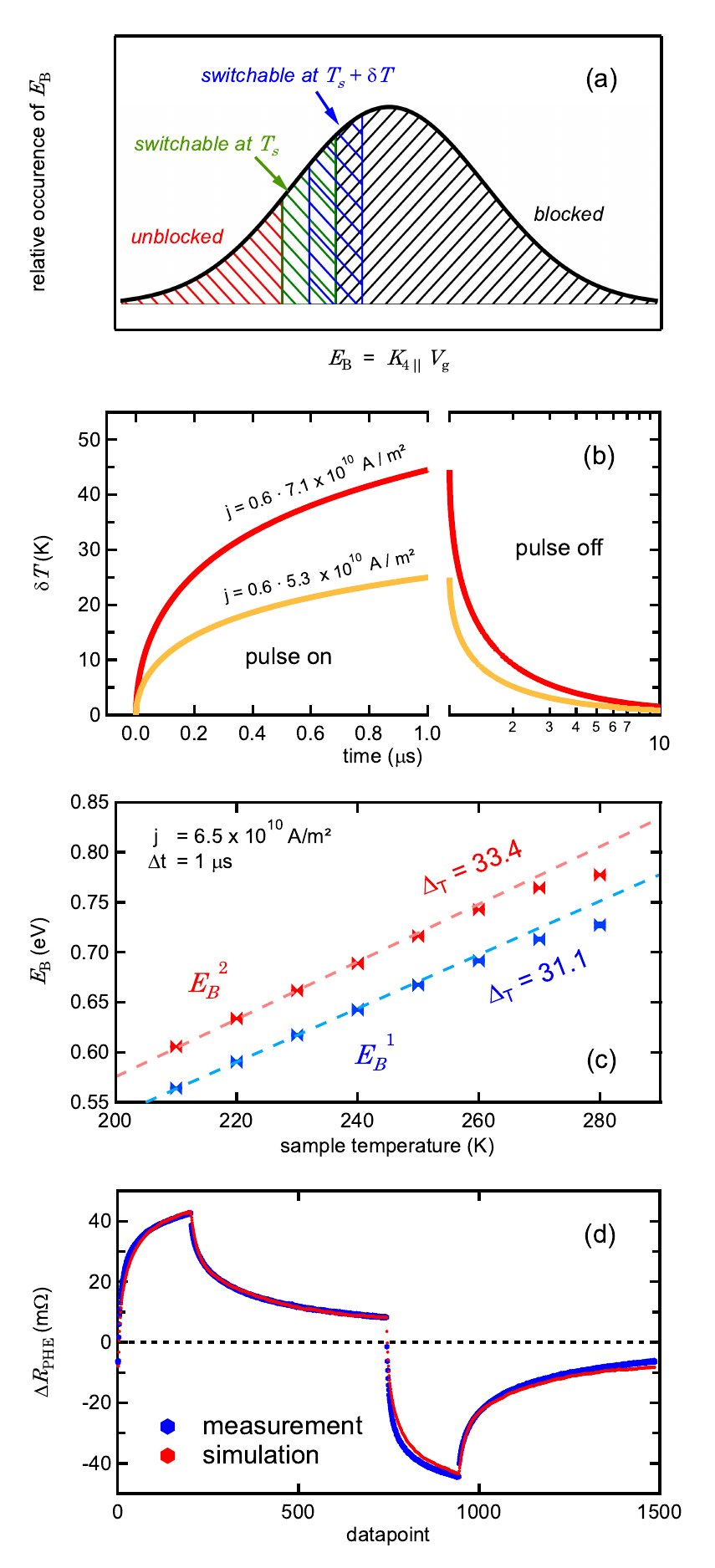}
	\caption[fig4]{\textbf{Model assumptions and simulation results.}
	(a)~Gaussian $E_\text{B} = K_{4\parallel} V_\text{g}$ distribution. The hatched areas indicate parts of the ensemble that are \textit{blocked} (red), \textit{switchable} (green/blue), or \textit{blocked}.
	(b)~Upper limit of the temperature rise $\delta T$ in the center region of the device caused by Joule heating during a $\Delta t = 1\,\text{\textmu s}$ pulse calculated for the investigated sample. The $x$-axis is split in a linear (pulse on) and a logarithmic (pulse off) part.
	(c)~$E_\text{B}^{1,2}$ calculated from $\tau_{1,2}$ using Eq.~\eqref{eq:EnergyBarrier}. The dashed lines are linear fits with no offset considering $T_\text{s}\le 240\,\text{K}$.
	(d)~Experiment (blue) to Monte Carlo simulation (red) comparison.
	}
	\label{fig:4}
\end{figure}

The model prediction regarding the dependence of $R_\text{e}$ on $T_\text{s}$ is confirmed by an Arrhenius plot shown in the inset of Fig.~\ref{fig:3}\,(a). As is indicated by the line fit in Fig.~\ref{fig:3}\,(b), our findings also confirm the expected dependence of $ R_\text{e} $ on the current density. To reason the local maximum in $\left| \Delta R_\text{a} \right|$ at $T_\text{s} = 260\,\text{K}$ we need a grain size distribution having a well defined maximum with $E_\text{B}^\text{center} = K_{4\parallel}\,V_\text{g}^\text{center} $. For simplicity we consider a Gaussian distribution, but our argument generally applies to any distribution. We define the thermal stability factor $\Delta_{T} = E_\mathrm{B} / k_\mathrm{B}T$ which governs the relaxation time of a grain via Eq.~\eqref{eq:NeelArrhenius}, $\tau = f_0^{-1} \exp(\Delta_T)$.  We categorize the grains roughly into \textit{unblocked} for $\Delta_T \lesssim 27$ ($\tau \lesssim 0.5\,\mathrm{s}$), \textit{switchable} for $27 \lesssim \Delta_T  \lesssim 44$, and \textit{blocked} if $\Delta_T \gtrsim 44$ ($\tau \gtrsim 10^7\,\mathrm{s}\approx 0.3\,\mathrm{yr}$), cf. Fig.~\ref{fig:4}\,(a). $ E_\text{B} = K_{4\parallel}\,V_\text{g} $ depends on the grain size distribution of our system, which is identical for all shown experiments. Joule heating results in a temperature increase $\delta T$ during a pulse allowing for the switching of a set of blocked grains, represented by the blue hatched area in Fig.~\ref{fig:4}\,(a). After the pulse, the temperature returns back to $T_\text{s}$ and during this cool-down the relaxation rate is enhanced. The measurement of $R_\text{PHE}$ is delayed by $2\,\text{s}$ after the burst. Thus, we measure grains which have been switched and did not relax back to equilibrium during this delay, which is effectively the case when $\tau \gtrsim 0.5\,\mathrm{s}$ during the pulse. The increased temperature $T_s + \delta T$ during the pulse shifts the subset of switchable grains to larger grain sizes. Thereby, larger grains become switchable, while small grains become unblocked during a pulse and occupy the four N\'{e}el-vector orientations uniformly. When the switchable subset during the pulse shifts through a maximum of the grain size distribution as a function of temperature, we therefore observe a maximum in $\left| \Delta R_\text{a} \right|$, while the efficiency $R_\text{e}$ is further increasing (cf. Fig.~\ref{fig:3}\,(a)).

In Fig.~\ref{fig:4}\,(b) we show $\delta T$ during and after a $1\,\text{\textmu s}$-pulse calculated with an analytical formula derived by You \textit{et al.} \cite{You2006} (cf. Appx.~\ref{Delta_T}). The current density in the center region of the star-like structure $j_\text{CR}$ is about 60\,\% of the nominal applied current density $j$ (cf. Appx.~\ref{curr_flow_sim}), thus, $j_\text{CR} = 0.6\,j$ is used for all quantitative investigations. After a steep increase of temperature during the first 200\,ns of a pulse, a weaker increase for long pulses is seen, where the curves scale with $j_\text{CR}^2$. After the pulse, the temperature drops rapidly within a few hundred nanoseconds and converges slowly to $\delta T = 0$ within tens of microseconds. The increase of $R_\text{e}$ and $\left| \Delta R_\text{a} \right|$ as a function of pulse width seen in Fig.~\ref{fig:3}\,(c) results from the dependence of $\delta T$ on the pulse width.

The $\tau_{1,2}$ dependence on $T_\text{s}$, $j$ and $\Delta t$ is shown in Fig.~\ref{fig:3}\,(d\,-\,f). While $\tau_{1,2}$ are independent on $j$ and $\Delta t$, they decrease with increasing $T_\text{s}$ and tend towards zero for $T_\text{s} > 280\,\text{K}$. We can calculate the energy barriers of the switchable grains using Eq.~\eqref{eq:EnergyBarrier}. The resulting $E_\text{B}^{1,2}( T_\text{s} )$ curves are shown in Fig.~\ref{fig:4}\,(c) and show linear dependencies on the temperature. By line fits we estimate the corresponding values of $\Delta_T$ to be 31.1 and 33.4 for the fast and the slow relaxation component, respectively. At lower temperature, the switchable subset of grains shifts to smaller grain volumes, such that $E_\mathrm{B} = k_\mathrm{B} T \Delta_T$ is in agreement with our categorization of switchable grains. From theory, the dominating intraband contribution to the NSOT is independent of the electron scattering time \cite{Zelezny2014}, and therefore, independent of resistivity and temperature. As a consequence, one expects that at lower temperature only grains with smaller energy barrier are switchable. Our results are fully compatible with this theoretical result. As mentioned earlier, the temperature increase during a pulse and the accompanied relaxation enhancement acts on a shorter time scale than our measurement can resolve. Formerly blocked grains that became switchable during the pulse and did not relax within the $2\,\text{s}$ delay are blocked again and, therefore, do not affect the relaxation time constants $\tau_{1,2}$. Hence, the observable resistance decay is determined by $T_\text{s}$ only.

Parameters that we extract from the experimental data using analytical solutions of our model can be inserted into full Monte Carlo simulations \cite{Meinert2018} to assess the internal consistency of the data analysis. In Fig.~\ref{fig:4}\,(d) we present one switching cycle of Fig.~\ref{fig:1}\,(e) and the corresponding simulation with optimized parameters. The simulation is done for $T_\text{s} = 260\,\text{K}$, $\Delta t = 5\,\text{\textmu s}$, and $j_\text{CR} = 0.6 \cdot 5.9 \times 10^{10}\,\text{A}/\text{m}^2$ considering three different grain volumes $V_\text{g}^{0,1,2}$ as suggested by our analysis. The respective energy barriers are $E_\text{B}^1 = 691\,\text{meV}$ and $E_\text{B}^2 = 743\,\text{meV}$ for the thermally unstable grains (cf.~Fig.~\ref{fig:4}\,(c)). $E_\text{B}^0$ accounts for the non-relaxing grains and, thus, cannot be determined by relaxation measurements. The temperature dependence of $R_\text{e}$, shown as Arrhenius plot in the inset of Fig.~\ref{fig:3}\,(a), allows to determine an activation energy $E_\text{A}$ for the switching that depends on each $E_\text{B}^{0,1,2}$. However, the evaluation of $E_\text{A}$ does not allow to draw conclusions for $E_\text{B}^{0}$ (cf. Appx.~\ref{arrhenius_eval}). We chose $E_\text{B}^0 = 2E_\text{B}^2 - E_\text{B}^1 = 795\,\text{meV}$ to be fixed in our simulations. Beyond these parameters, the simulation uses literature values for $\chi = 3\,\text{mT}/(10^{11}\text{A}/\text{m}^2)$~\cite{Wadley2016} and $L = 7\mu_\text{B}$~\cite{Wadley2015}. $A$, $K_{4\parallel}$, and the population of each grain size are free parameters. With a biaxial anisotropy energy density of $K_{4\parallel}= 1.2\,\text{\textmu eV}/\text{(unit cell)} \approx 2.1\,\text{kJ}/\text{m}^3$ a good match between simulation and experiment is achieved. The corresponding grain volumes are of the order $V_\text{g} \approx (37 \dots 39\,\text{nm})^3$. This fairly large value $V_\text{g}$ indicates that our film may have magnetic domains that span several grains, in which case the magnetic domain volume plays the role of $V_\mathrm{g}$ in our formalism. Antiferromagnetic domains can form across grain boundaries in antiferromagnets, e.g. due to magnetoelastic coupling which allows that the N\'{e}el-order can couple via mutual mechanical stress and magnetostriction of neighboring grains \cite{Gomonay2002}.

\section{Conclusion}

In summary, we demonstrated that NSOT switching is possible in magnetron sputter-deposited \mbox{CuMnAs} thin films that are easy to grow, paving the way for a broader scientific community to study this material. In our electrical experiments, the sample temperature $T_\text{s}$, current density $j$, and pulse width $\Delta t$ were varied and the observed dependencies are discussed using a macroscopic thermal-activation model. We demonstrate that the switching and relaxation can be simulated with quantitative agreement by our model using values obtained from the experiment and theory so that only three free parameters are left. Our study presents a quantitative analysis scheme for the observed switching, which will aid the rational optimization of the film growth and thereby opens an avenue for room-temperature application of the NSOT switching.

\subsection*{Acknowledgments}

The authors thank K. Rott and D. Ramermann for the transmission electron microscopy images.

\appendix

\counterwithout{equation}{section}

\setcounter{table}{0}
\renewcommand{\thetable}{A\arabic{table}}%
\setcounter{figure}{0}
\renewcommand{\thefigure}{A\arabic{figure}}%
\setcounter{equation}{0}
\renewcommand{\theequation}{A\arabic{equation}}

\section{Sample preparation and characterization}\label{Properties}

\begin{figure}
	\centering
	\includegraphics{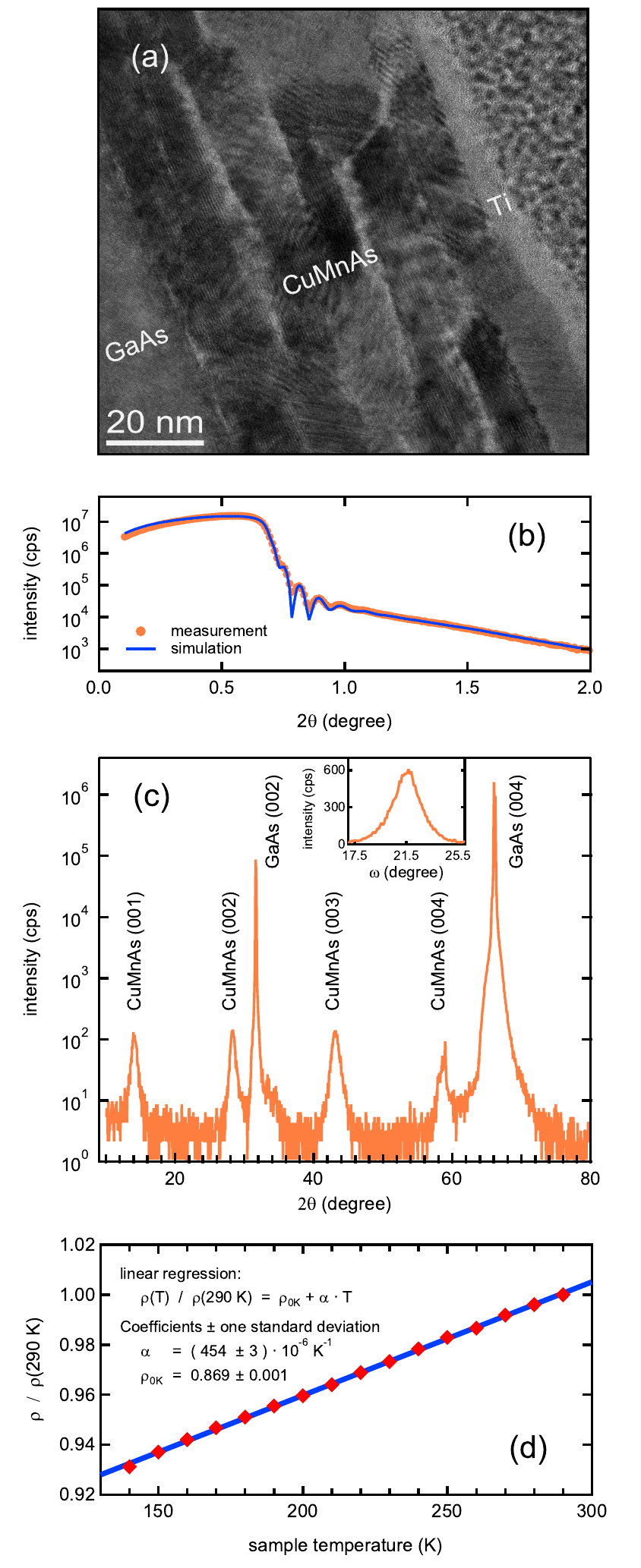}
	\caption[figS1]{\textbf{Properties of the sample.}
		(a)~XRR and (b)~XRD measurements on the GaAs~/ \mbox{Cu}\mbox{Mn}\mbox{As}~68~/ Ti~5 stack. The inset in (b) shows the rocking curve of the (003) peak.
		(c)~TEM image of the sample.
		(d)~Temperature dependent change of the resistivity normalized to the resistivity at $290\,$K sample temperature with a linear regression applied to the data.
		}
	\label{fig:S1}
\end{figure}

We apply dc~magnetron sputtering to deposit the \mbox{CuMnAs} films from a composite target with Cu$_{0.3}$Mn$_{0.3}$As$_{0.4}$ stoichiometry on HCl-etched GaAs~(001) substrates at a deposition temperature of $410\,^\circ\text{C}$. A capping layer of Ti was deposited after cool-down of the sample. We designed the deposition to grow a GaAs~/ \mbox{Cu}\mbox{Mn}\mbox{As}~80\,\text{nm}~/ Ti~5\text{nm} stack taking the growth rate determined from a reference sample deposited at room temperature. However, the actual \mbox{CuMnAs} film thickness reads $68\,\text{nm}$, verified by transmission electron microscopy (TEM) and x-ray reflectivity (XRR) measurements (cf. Fig.~\ref{fig:S1}\,(a) \& (b)). The TEM lamella was cut parallel to the (110) direction of the \mbox{GaAs} substrate. Energy-dispersive x-ray spectroscopy (EDX) shows the presence of oxygen throughout the Ti film and in the bright areas of the CuMnAs layer. Since CuMnAs likely oxidizes under ambient conditions we relate this oxygen accumulation to our TEM preparation process which includes contact to air. Nevertheless, the oxidation is limited to certain areas of the film, indicating that the crystal is more vulnerable to oxygen in these regions. We interpret these weak spots as grain boundaries. The XRR curve is fitted using the GenX reflectivity fitting package \cite{Bjoerck2007_GenX}. The film roughness of $4.6 \dots 5.0\,\text{nm}$ obtained by the XRR fit is consistent with atomic force microscopy measurements that state $R_\text{rms} = 4.8\,\text{nm}$ root-mean-square roughness and $R_\text{p2p} = 39\,\text{nm}$ peak-to-peak roughness. The $5\,\text{nm}$ Ti capping layer preserves the film from bulk oxidation for at least several months, while an uncapped layer completely oxidizes within weeks. In Fig.~\ref{fig:S1}\,(c) x-ray diffraction (XRD) data is shown. The spectrum verifies the crystallization of the CuMnAs with an out-of-plane lattice parameter $c=6.286\,\text{\AA}$, which is $0.5\,\%$ smaller than reported for samples prepared with molecular beam epitaxy (MBE)\cite{Wadley2013}. The full-width at half-maximum (FWHM) of the (003) peak is $1.02\,^\circ$ and the FWHM of its rocking curve is $2.43\,^\circ$. The out-of-plane crystallite size is therefore in the order of $10\,\text{nm}$ following the Scherrer equation. This is consistent with the TEM imaging where the in-plane dimension of the crystallites appears to be larger. From our simulation result for $V_\text{g} \approx (37 \dots 39\,\text{nm})^3$ we find the in-plane crystallite diameter to be in the order of $80\,\text{nm}$ in the case of one antiferromagnetic domain per crystallite. Due to the rough surface, we can only estimate $\rho \approx 430\,\text{\textmu}\Omega\,\text{cm}$ at room temperature using four-point-measurements and the thickness subtracted by $R_\text{p2p} / 2$. Its temperature dependence is plotted in Fig.~\ref{fig:S1}\,(d) as $\rho(T) / \rho(T=290\,\text{K})$ as a function of the sample temperature $T$ within the investigated temperature range. The resistivity drops by $7\,\%$ when decreasing the temperature from 290\,K to 140\,K. From a linear regression we extrapolate that the residual resistance is $\rho(0\,\text{K}) = 0.869\,\rho(T=290\,\text{K})$. Two neighboring devices on the same sample were used for this investigation. The $j$ and $\Delta t$ variations were performed on the same device, whereas the $T_\text{s}$ variation was done on the second one.

\section{Residual resistance gap}\label{gap}

\begin{figure}
	\centering
	\includegraphics{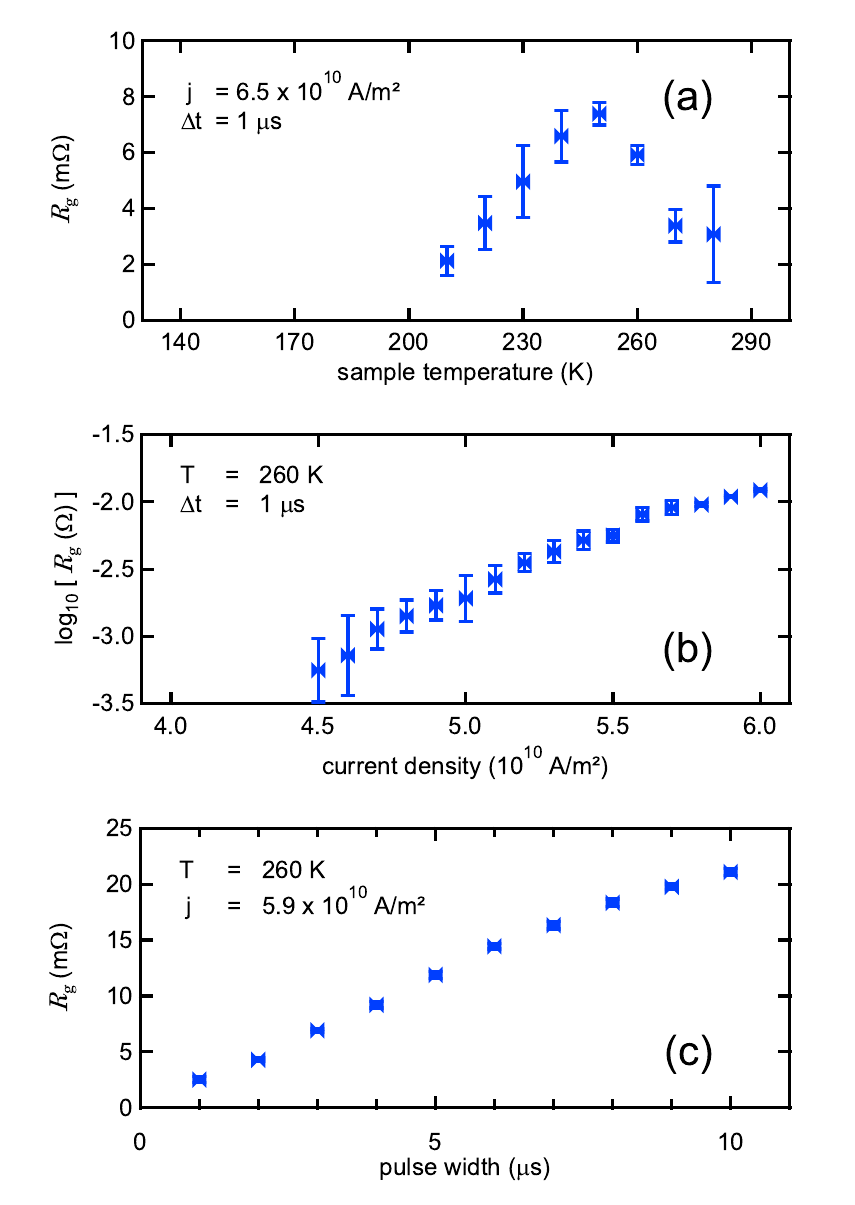}
	\caption[figS2]{\textbf{The residual resistance gap.}
		Dependence of $R_\text{g}$ on (a) $T_\text{s}$, (b) $j$ and (c) $\Delta t$. The $x$-axes are chosen to match the corresponding plots in Fig.~\ref{fig:3}\,(a-c).
		}
	\label{fig:S2}
\end{figure}

The fitting routine that extracts $R_\text{e}$ from our data also provides values for the residual resistance $d_0$, which  represents switched grains with long relaxation times. We define the difference of $d_0$ after pulsing along $x$ and $y$ direction as residual resistance gap 
\begin{align}
R_\text{g} = \left| d_0^{\parallel x} - d_0^{\parallel y} \right|
\end{align}
which is a measure for the resistance difference between two electrically set antiferromagnetic states that are long-term stable and, thus, can be used for information storage. For measurements of 10 switching cycles we calculate $R_\text{g}$ five times for subsequent cycles without any doubling. The standard deviation is taken as error. The dependence of $R_\text{g}$ on $T_\text{s}$, $j$, and $\Delta t$ is presented in Fig.~\ref{fig:S2}\,(a),\,(b), and (c), respectively. The curves are similar to the $\left| \Delta R_\text{a} \right|$ dependence for all varied quantities (cf. Fig.~\ref{fig:3}\,(a-c) in the main text). We therefore conclude, that the value of $\left| \Delta R_\text{a} \right|$ is strongly linked to the amount of grains that are usually blocked at $T_\text{s}$ but switchable during the pulse. Thus, these grains are thermally stable after the pulse.

\section{Derivation of the planar Hall resistance decay}\label{rate}

In general, the N\'eel vector $\bm{L}$ can be in one of $p$ different states if we assume $p$-fold rotational symmetry. Such a system can be described as
\begin{align}
\bm{N} = \left(\begin{array}{c} N_1 \\ N_2 \\ \vdots \\ N_p \end{array}\right)
\end{align}
with $N_i$ being the relative occupation number of each state and $\sum_i N_i = 1$. Thermal activation allows $\bm{L}$ to hop randomly from one state to another. Without electrical current, all states are energetically degenerate and, thus, the change of $N_i$ depends on its occupation only. Therefore we can write
\begin{align}\label{eq:DGL}
\frac{dN_i}{dt} = \sum_{j=1}^{p} \left[ \left( \frac{1}{p} - \delta_{ij} \right) \nu N_j \right]
\end{align}
as the time derivative of $N_i$. $\delta_{ij}$ is the Kronecker delta and the thermal activation rate $\nu = 1/\tau$ is given by the N\'{e}el-Arrhenius equation (Eq.~\eqref{eq:NeelArrhenius} in the main text). Restricting ourselves to biaxial anisotropy ($p=4$), Eq.~\eqref{eq:DGL} can be written in matrix form as
\begin{align}\label{eq:RateEquation}
\frac{d\bm{N}}{dt} =
\frac{\nu}{4}
\left(\begin{array}{cccc}
-3 & 1 & 1 & 1 \\ 
1 & -3 & 1 & 1 \\ 
1 & 1 & -3 & 1 \\ 
1 & 1 & 1 & -3
\end{array}\right)
\left(\begin{array}{c} N_1 \\ N_2 \\ N_3 \\ N_4 \end{array}\right)
\end{align}
which has the general solution
\begin{align}
\bm{N}(t) &= e^{ - \frac{t}{\tau} }
\left[
C_1 \left(\begin{array}{c} -1 \\ 0 \\ 0 \\ 1 \end{array}\right)
+ C_2 \left(\begin{array}{c} -1 \\ 0 \\ 1 \\ 0 \end{array}\right)
+ C_3 \left(\begin{array}{c} -1 \\ 1 \\ 0 \\ 0 \end{array}\right)
\right] \nonumber \\
&+ \frac{1}{4}\left(\begin{array}{c} 1 \\ 1 \\ 1 \\ 1 \end{array}\right)\text{.}
\end{align}
$N_{1,3}$ and $N_{2,4}$ account for $\bm{L}$ being (anti)parallel to the $x$- or $y$-axes of our experiment, respectively (cf.~Fig.~\ref{fig:1}\,(a) in the main text). Assuming a burst applied at $t=0$ switches all grains so that $N_1 = 1$, one obtains the coefficients $C_1 = C_2 = C_3 = -1/4$. The measured planar Hall resistance is calculated as
\begin{align}
\hspace{-1mm}R_\text{PHE}(t) &= A_\text{AMR}\,\left[ N_1(t) + N_3(t) - N_2(t) - N_4(t) \right] \\
&= A_\text{AMR}\,\exp\left( - \frac{t}{\tau} \right),
\end{align}
where $A_\text{AMR}$ is the AMR amplitude. Since in our experiment we do not create a state with unidirectional $\bm{L}$ at $t=0$, $A_\text{AMR}$ is replaced by $R_\text{PHE}(t=0) < A_\text{AMR}$ in our analysis (cf.~Eq.~\eqref{eq:RelaxationModel} in the main text).

\section{Calculation of $\delta T$}\label{Delta_T}

To estimate the temperature rise during a pulse we apply a model derived by You \textit{et al.} which states that\cite{You2006}
\begin{align}\label{eq:TemperatureRise}
\delta T (t) &= \frac{whj_\text{CR}^2}{\pi \kappa_S \sigma} \Bigg[ \text{arcsinh}\left( \frac{2 \sqrt{\mu_S t\, }}{\alpha w} \right) \nonumber\\
&+ \theta\left( t-\Delta t \right) \text{arcsinh}\left( \frac{2 \sqrt{\mu_S \left(t - \Delta t \right)}}{\alpha w} \right) \Bigg]
\end{align}
with the film thickness $h$, the currentline width $w$, the electrical conductivity $\sigma$ and the substrate parameters $\kappa_S$ and $\mu_S$, namely the heat conductivity and the thermal diffusivity. $\alpha = 0.5$ is chosen as suggested by You \textit{et al.}. The model assumes a two-dimensional heat flow and we apply a constant voltage to generate our pulses. The temperature rise of the device will reduce $\sigma$ and, thus, $j$ will fall during the pulse. Therefore the calculation yields an upper estimate for $\delta T$ in this experiment. This equation  is used in our Monte Carlo simulation.

\section{Current flow distribution}\label{curr_flow_sim}

\begin{figure}
	\centering
	\includegraphics{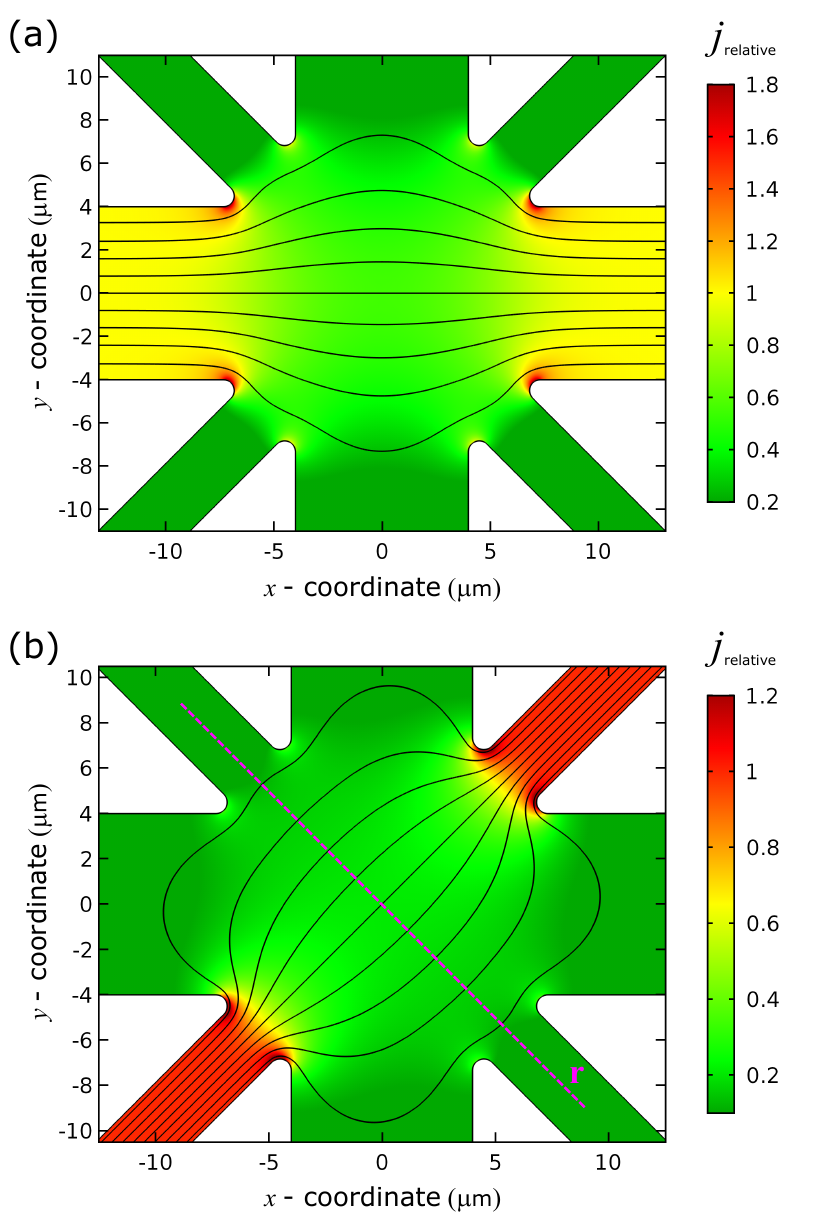}
	\caption[figS3]{\textbf{Current flow distribution.}
		Finite-element simulation of the relative current flow distribution in our device.
		(a) For a constant voltage being applied at the pulse line in $x$-direction.
		(b) For a constant voltage being applied at the probe line in $(x,y)$-direction. The purple line points in $(-x,-y)$-direction.
		}
	\label{fig:S3}
\end{figure}

We performed finite-element simulations of the current flow distribution in our star-like structures for the switching pulse and the probe current, shown in Fig.~\ref{fig:S3} (a) and (b), respectively. Similar to simulations for 4-arm Hall crosses \cite{Meinert2018} we find an inhomogeneous current flow with hot spots at the corners. For pulsing the current density in the center of the cross reads 55\% of the current density in the leads while we find up to 85\% higher current densities in the hot spots. However, compared to the 4-arm crosses the area of nearly homogeneous current density is quite large. The switching is evaluated by PHE measurements
\begin{align}\label{eq:Rphe}
R_\text{PHE} = \frac{V_\perp}{I_\text{probe}}
\end{align}
with $I_\text{probe}$ being the current through one probe line and $V_\perp$ being the voltage measured at the perpendicular probe line. The current density distribution of the switching pulse and, thus, the resulting magnetic configuration, are inhomogeneous in our experiment. To illustrate the influence of the inhomogeneous magnetic configuration we describe $V_\perp$ by considering the line integral
\begin{align}\label{eq:Vperp1}
V_\perp = \int_{\bm{r}} \bm{E}(\bm{j}_\text{probe}) \cdot \text{d}\bm{r}
\end{align}
along the curve $\bm{r}$, visualized in Fig.~\ref{fig:S3}\,(b). Although the presence of an AMR disturbs the current density profile slightly, the condition $\bm{j}_\text{probe} \perp \text{d}\bm{r}$ is approximately fulfilled and Eq.~\eqref{eq:Vperp1} can be written as
\begin{align}\label{eq:Vperp2}
V_\perp   = h \, \int_{\bm{r}} \text{d}r \, j_\text{probe} \, A_\text{AMR} \, \sin 2\theta
\end{align}
with $\theta = \angle \left( \bm{j}_\text{probe},\bm{L} \right)$ and $j_\text{probe} \, A_\text{AMR} \, \sin 2\theta = E_\parallel  / \, h$ being the electrical field component parallel to $\text{d}\bm{r}$ divided by the film thickness $h$. 

From this considerations we see that a large portion of $V_\perp$ originates from the central region of the star-like structure. Hence, the PHE measurement is mostly sensitive on the switching in this region and therefore we introduce the center region~(CR) current density
\begin{align}
j_\text{CR} = 0.6 \, j
\end{align}
for our quantitative analysis.

\section{Arrhenius plot evaluation}\label{arrhenius_eval}

\begin{figure}
	\centering
	\includegraphics{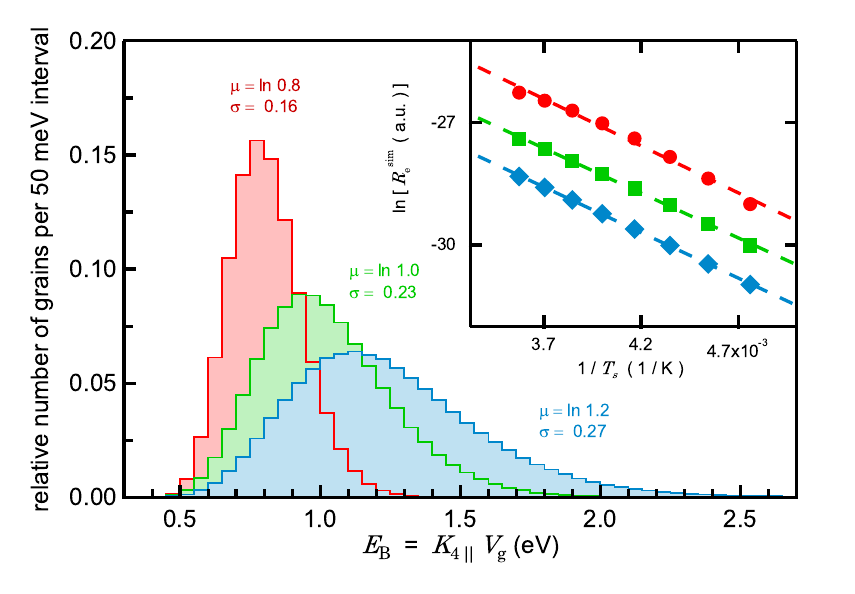}
	\caption[figS4]{\textbf{Different grain size distributions.}
		Histograms for ensembles of $10^6$ grains each where $E_\text{B} = K_{4\parallel}V_\text{g}$ is lognormal distributed. The inset shows the Arrhenius plot for each distribution in the respective color.
	}		
	\label{fig:S4}
\end{figure}

In the Discussion we identified $\left| \Delta R_\text{PHE}^\text{burst} \right| \approx R_\text{e}$. Using this identity and taking the natural logarithm of Eq.~\eqref{eq:BurstApproximation} in the main text we obtain
\begin{align}
\ln R_\text{e} \approx \ln \left(\frac{Q A f_0}{j_\text{CR} w h}\right) + \left( \frac{L \chi V_\text{g} j_\text{CR}}{\sqrt{2} k_\text{B} V_\text{c}} - \frac{E_\text{B}}{k_\text{B}}\right) \frac{1}{T}
\end{align}
applying $j_\text{CR}$ instead of $j$. Thus, the results of the line fit $s_1 - s_2 / T_\text{s}$ evaluated in the Arrhenius plot, shown in the inset of Fig.~\ref{fig:3}\,(a) in the main text, gives access to the reduced activation energy
\begin{align}\label{eq:EffectiveActivationEnergy}
E_\text{A}^\text{red} &= k_\text{B} s_2 \\ &=  E_\text{B} - \frac{L \chi V_\text{g} j_\text{CR}}{\sqrt{2} V_\text{c}} = \left( 195 \pm 8 \right) \, \text{meV}
\end{align}
of the switching. Note that the Arrhenius plot evaluates $ E_\text{A}^\text{red} $ with respect to $T_\text{s}$ although the switching happens with a time dependent temperature $T_\text{s} < T < T_\text{s} + \delta T$. Hence, $ E_\text{A}^\text{red} < E_\text{A}$ is underestimated. The error is obtained from the standard deviation of the fit. In the following we discuss, why no further quantitative conclusion can be drawn from the analysis of the Arrhenius plot. In Fig.~\ref{fig:S4} we show histograms of three ensembles with $10^6$ grains each, where the distribution of $E_\text{B} = K_{4\parallel}V_\text{g}$ is lognormal, i.e. $\mathcal{LN}(\mu,\sigma)$. $\mu$ is the mean value and $\sigma$ the standard deviation of the distribution.

From a known distribution we can calculate the switching efficiency
\begin{align}\label{eq:R_simulated}
R_\text{e}^\text{sim} \propto \sum_{i=1}^{10^6} \Bigg(  &l_\text{g}^i 
\exp \left[  \left( \xi\,V_\text{g}^i - \frac{K_{4\parallel}}{k_\text{B}}\,V_\text{g}^i \right) \frac{1}{T} \right] \cdot \nonumber \\
~&\exp\left[ - 2\,\text{s} \cdot f_0 \exp \left( - \frac{K_{4\parallel}\,V_\text{g}^i}{k_\text{B} T} \right) \right]
\Bigg)
\end{align}
with $\xi = L \chi j_\text{CR} / (\sqrt{2} k_\text{B} V_\text{c})$ following Eq.~\eqref{eq:BurstApproximation} in the main text. We use $K_{4\parallel} V_\text{c} = 1.2\,\text{\textmu eV}$ ($K_{4\parallel} \approx 2.1\,\text{kJ}/\text{m}^3$). The second exponential function accounts for the relaxation during the delay of $2\,\text{s}$ in our experiment (cf. Eq.~\eqref{eq:NeelArrhenius} in the main text). Each summand contributes to the measured transverse voltage in proportion to the edge length $l_\text{g} \approx \sqrt[3]{V_\text{g}}$ of the respective grain. The Arrhenius plot of $R_\text{e}^\text{sim}$ is shown in the inset of Fig.~\ref{fig:S4} for three different grain size distributions. $\mathcal{LN}_1(\ln 0.8,0.16)$, $\mathcal{LN}_2(\ln 1.0,0.23)$, and $\mathcal{LN}_3(\ln 1.2,0.27)$ are represented by similar $E_\text{A}^\text{red} = \{198 \pm 8,~189 \pm 4,~192 \pm 2\}\,\text{meV}$ although their shapes are diverse. Hence, $E_\text{A}^\text{red}$ does not allow to draw conclusions about the underlying grain size distribution and, thus, the Arrhenius plot evaluation only yields an effective quantity.

\end{document}